\documentclass{PoS}

\pdfoutput=1

\usepackage{amsmath,amssymb}
\graphicspath{{./graphs/}}

\newcommand\emailUsername[2]{{\tt\href{mailto:#1@#2}{#1}}}
\newcommand\emailFullpath[2]{{\tt\href{mailto:#1@#2}{#1@#2}}}

\title{Thermal dilepton rates from quenched lattice QCD}

\ShortTitle{Thermal dilepton rates from quenched lattice QCD}

\author{O. Kaczmarek, E. Laermann, \speaker{M. M\"uller}\\
Fakult\"at f\"ur Physik, Universit\"at Bielefeld, D-33615 Bielefeld, Germany\\
E-mail: 
\emailUsername{okacz}{physik.uni-bielefeld.de},
\emailUsername{edwin}{physik.uni-bielefeld.de},
\emailFullpath{mmueller}{physik.uni-bielefeld.de}
}

\author{F. Karsch\\
Fakult\"at f\"ur Physik, Universit\"at Bielefeld, D-33615 Bielefeld, Germany\\
Physics Department, Brookhaven National Laboratory, Upton, NY 11973, USA\\
E-mail: \emailFullpath{karsch}{quark.phy.bnl.gov}
}

\author{H.-T. Ding, S. Mukherjee\\
Physics Department, Brookhaven National Laboratory, Upton, NY 11973, USA\\
E-mail:
\emailUsername{htding}{quark.phy.bnl.gov},
\emailFullpath{swagato}{quark.phy.bnl.gov}
} 

\author{A. Francis\\
Institut f\"ur Kernphysik, Johannes Gutenberg-Universit\"at Mainz,
D-55099 Mainz, Germany\\
E-mail: \email{francis@kph.uni-mainz.de}
}

\author{W. Soeldner\\
Fakult\"at f\"ur Physik, Universit\"at Regensburg, D-93053 Regensburg, Germany\\
E-mail: \email{wolfgang.soeldner@physik.uni-regensburg.de}
}

\abstract{
We present new lattice results on the continuum extrapolation of the 
vector current correlation function. Lattice calculations have 
been carried out in the deconfined phase at a temperature of  
$T = 1.1 T_c$, extending our previous results at $T = 1.45 T_c$,
 utilizing quenched non-perturbatively 
clover-improved Wilson fermions and light quark masses.
A systematic analysis on multiple lattice spacings allows to 
perform the continuum limit of the correlation function and 
to extract spectral properties in the continuum limit.

Our current analysis suggests the
results for the electrical conductivity  
are proportional to the temperature
and the thermal dilepton rates 
in the quark gluon plasma 
are comparable for both temperatures.
Preliminary results of the 
continuum extrapolated correlation function
at finite momenta, which relates to thermal 
photon rates, are also presented.}

\FullConference{Xth Quark Confinement and the Hadron Spectrum,\\
		October 8-12, 2012\\
		TUM Campus Garching, Munich, Germany}

\begin{document}

\section{Introduction}

Current Heavy-Ion-experiments at LHC and RHIC reach energies 
in a region where the transition from normal nuclear matter 
to a quark gluon plasma occurs. In this phase thermally produced 
dileptons and photons are important experimental
observables. To understand their yields, hydrodynamic models
for the evolution of the medium need detailed
knowledge of the dilepton and photon rates as
well as the transport coefficients during
the evolution of the system \cite{Rapp:2009yu}. 
These require non-perturbative ab initio lattice QCD 
calculations as input.

On the lattice, hadronic correlation functions for 
different particle channels can be calculated at zero
and at fixed finite temperatures. 
From the vector spectral function the electrical
conductivity and the dilepton rates of the medium 
can be extracted.
The spectral function has recently been 
extracted 
by employing phenomenologically inspired ans\"atze
\cite{pos-olaf, pos-hengtong}.

Our previous work on dilepton rates
provided the first results for a continuum extrapolation
of the vector meson correlation at a temperature of 
$T/T_c=1.45$ and light quark masses \cite{Ding:2010ga}.
An ansatz for the spectral function could successfully 
be used to describe this dataset, thereby the dilepton rate could
be obtained at a first relevant temperature.
Together with results for the electrical conductivity
at different temperatures $1.16 < T/T_c < 2.98$ \cite{Francis:2011bt}
this motivated the systematic study of the temperature
dependence of the vector spectral function on the lattice.
For results using two dynamical flavors at finite lattice
spacings see \cite{Brandt:2012jc}.

The work presented here extends the continuum extrapolated
quenched calculations to a second relevant temperature $T/T_c=1.1$.

\section{Thermal vector correlator and spectral function}

\subsection{Vector correlation function}

The Euclidean time two-point correlation function $G(\tau,\vec p)$ 
of the vector current $J_\mu$
is a quantity directly accessible in lattice QCD calculations,

\begin{equation}
G_{\mu\nu}(\tau,\vec p) = \int d^3x  J_\mu(\tau,\vec x) J_\nu^\dag(0,\vec0)  e^{i\vec p \vec x} 
 \quad \text{with} \quad J_\mu(\tau,\vec x) = \bar q(\tau, \vec x) \gamma_\mu q(\tau,\vec x).
\end{equation}

Only contributions of quark line connected diagrams are included. Disconnected diagrams
cause a high numerical effort, but are expected to be small in the high temperature phase of QCD
\cite{Allton:2005gk, Gavai:2001ie}.
The correlation function
directly relates to the spectral function via 

\begin{equation}
G_H(\tau, \vec p,T) = \int_0^\infty \frac{d\omega}{2\pi} \rho_H(\omega, \vec p, T) 
\frac{\cosh(\omega(\tau - 1/2 T))}{\sinh{(\omega/2T)}} \quad \text{with:} \quad H=00, ii, V.
\label{spfToCorr}
\end{equation}

Here $\rho_{ii}$ denotes a sum over the spatial components, while $\rho_{00}$ denotes
the time-like components. The full vector spectral function 
is denoted by $\rho_V = \rho_{00} + \rho_{ii}$. As lattice QCD calculations are
carried out at a fixed temperature, the explicit temperature dependence of $\rho$ 
is dropped and observables are usually given in units of 
the temperature $T$.

\subsection{Ansatz for the spectral function}

The time-like component of the vector correlator $G_{00}$ and thereby the corresponding
transformation of the spectral function $\rho_{00}$ is related to 
the quark number susceptibility $\chi_q$. Since the quark 
number is conserved, the correlator is constant in 
(here Euclidean) time, $G_{00}(\tau T) = - \chi_q T$
and its spectral representation is given by a delta function

\begin{equation}
\rho_{00}(\omega) = -2 \pi \chi_q \omega \delta(\omega).
\label{chiEquation}
\end{equation}

The spatial components of the spectral functions increase quadratically for large
values of $\omega$, in the free field limit for massless quarks to

\begin{equation}
\rho_{ii}^\text{free}(\omega) = 2 \pi T^2 \omega \delta(\omega) + \frac{3}{2\pi} \omega^2 \tanh(\omega/4T).
\end{equation}

In this limit, the delta peak in the spatial and the time-like component
of the spectral function cancel. In the interacting theory however, this is
not the case: The time-like component maintains a delta peak since
it is linked to the conserved current. In the spatial component however
the delta peak is smeared out and expected to be described by
a Breit-Wigner peak \cite{Aarts:2002cc,Moore:2006qn,Petreczky:2005nh, Hong:2010at} 

\begin{equation}
\rho_{ii}^\text{interac.}(\omega) = \chi_q c_{\text{BW}} \frac{\omega \Gamma}{\omega^2 + (\Gamma/2)^2}
 + (1+\kappa) \frac{3}{2\pi} \omega^2 \tanh(\omega/4T).
\label{spfAnsatz}
\end{equation}

This ansatz leaves three parameters, the strength ($c_{\text{BW}}$) and width ($\Gamma$) of
the Breit-Wigner peak as well as $\kappa$, which accounts for the deviation from free theory.
The relation of this ansatz to the correlator obtained on the lattice  
is given by \eqref{spfToCorr}.

The fits are not performed directly to the correlation function $G_{ii}$ 
but to a set of two ratios: The correlation function is
normalized by the quark number susceptibility (as given in 
\eqref{chiEquation}), resulting in
a dimensionless quantity independent of
renormalization constants. It is also normalized by the free field
correlation function $G_V^\text{free}(\tau T)$, yielding
a smooth function that does not fall off over multiple decades
like the correlation function. 
Furthermore, due to asymptotic freedom, the correlation function
approaches the non-interacting limit at asymptotically small distances.
The spectral function is thereby fitted to reproduce

\begin{equation}
\frac{ G_{ii}(\tau T) /  G_{00} }{ 
G_V^\text{free}(\tau T) / G_{00}^\text{free}  }.  
\label{fitRatios}
\end{equation}

Having obtained the spectral function, relevant properties
of the medium can be calculated, e.g. the electrical conductivity as

\begin{equation}
\frac{\sigma}{T} = \frac{C_\text{em}}{6} \lim_{\omega\rightarrow 0} \frac{\rho_{ii}(\omega)}{\omega T}
\quad \rightarrow \quad \sigma(T) / C_\text{em} = 2 \chi_q c_\text{BW} / (3 \Gamma)
\label{elcon}
\end{equation}
where $C_\text{em}$ is given by the
elementary charges $Q$ of the quark flavor $f$ 
as $C_\text{em}=\sum_f Q_f^2 $, 
and the thermal production rate of dilepton pairs as

\begin{equation}
\frac{dW}{d\omega d^3 p} = \frac{5 \alpha^2}{54 \pi^3 } 
\frac{1}{\omega^2 ( e^{\omega / T} - 1 )} 
\label{dileptonrate}
\rho_{ii}(\omega,p,T).  
\end{equation}

\subsection{Thermal moments}

Using the ansatz above to perform a three-parameter fit
of the spectral function, the Breit-Wigner parameters
are most sensitive to the low $\omega$ region, thereby
to the large distance behavior of the correlator.

This region of the correlator can be further constrained by
calculating the curvature of the correlator for large Euclidean
distances $\tau T$. The thermal moments

\begin{equation}
G_H^{(n)} = \frac{1}{n!} \left. \frac{d^n G_H(\tau T)}{d(\tau T)^n} 
\right\vert_{\tau T=1/2}  = 
\frac{1}{n!} \int_0^\infty \frac{d\omega}{2 \pi}
\left ( \frac{\omega}{T} \right )^n 
\frac{\rho_H(\omega)}{\sinh(\omega/2T)} 
\quad\text{with}\quad H=ii,V
\end{equation}
are given as the Taylor coefficients of the correlation
function expanded around the midpoint

\begin{equation}
G_H(\tau T) = \sum_{n=0}^{\infty} G_H^{(2n)} ( \frac{1}{2} -\tau T)^{2n}.
\end{equation}

For the infinite temperature, free field limit, the thermal moments
can be calculated analytically. For the analysis,
the ratios of interacting to free midpoint subtracted correlation
functions, 

\begin{equation}
\Delta_V(\tau T) = 
\frac{ G_V(\tau T) - G^{(0)}_V }{ G_V^\text{free}(\tau T) - G^{(0),\text{free}}_V }
=\frac{G_V^{(2)}}{G_V^{(2),\text{free}}}
\left (1 + ( R_V^{(4,2)} - R_{V,\text{free}}^{(4,2)}  ) (\frac{1}{2} - \tau T)^2  + \dots \right )
\label{midpoint-sub-corr}
,
\end{equation}

have been calculated, where $R_V^{(n,m)} = G_V^{(n)} / G_V^{(m)}$. The first two terms
(up to quadratic) are then used in the fit.

For the $1.45 T_c$ dataset, restricting the spectral function fit 
to also reproduce these first two thermal moments allowed to
obtain a stable fit. 
Since the new $1.1 T_c$ dataset allows for a more precise
continuum extrapolation down to a distance
$\tau T_{\text{min}}=0.15$,
which is smaller than the distance 
$\tau T_{\text{min}}=0.25$ reachable at $1.45 T_c$,
we find
that stable fits at this temperature 
(giving results 
within the systematic errors of this analysis) 
can also be performed
without including the thermal moments as an additional constraint.

\section{Results for correlators and spectral function fits}

\begin{table}

\begin{center}
\begin{tabular}{|c|c|c|c|c|c|c|}
\hline
$N_\tau$ & $N_\sigma$ & $\beta$ & $\kappa$ & $ 1/a [\text{GeV}]$ & $  a [\text{fm}]$ & \#conf \\
\hline
32 & 96 & 7.192  & 0.13440 &  9.65 & 0.020 & 314 \\ 
48 & 144 & 7.544 & 0.13383 & 14.21 & 0.015 & 315 \\
64 & 192 & 7.793 & 0.13345 & 19.30 & 0.010 & 242 \\ 
\hline
\end{tabular}
\end{center}

\caption{Summary of simulation parameters}
\label{simulationParameters}
\end{table}

Vector correlators are obtained 
on lattices of size 
$N_\sigma^3 \times N_\tau$, where
for the
new dataset at $1.1 T_c$ 
three fixed lattice spacings $(aT)^{-1}=N_t=32,48,64$
where chosen for the continuum extrapolation.
For each size $\beta$ was set to match the fixed
temperature and $\kappa$ was tuned to
light quark masses (see table \ref{simulationParameters}).
From the 
previous study at $1.45 T_c$, finite volume
effects are found to be small 
for a fixed aspect ratio $N_\sigma / N_\tau = 3$,
so the lattice sizes were set to
$96^3 \times 32$ , $144^3 \times 48$
and $192^3 \times 64$.

\begin{figure}
\includegraphics{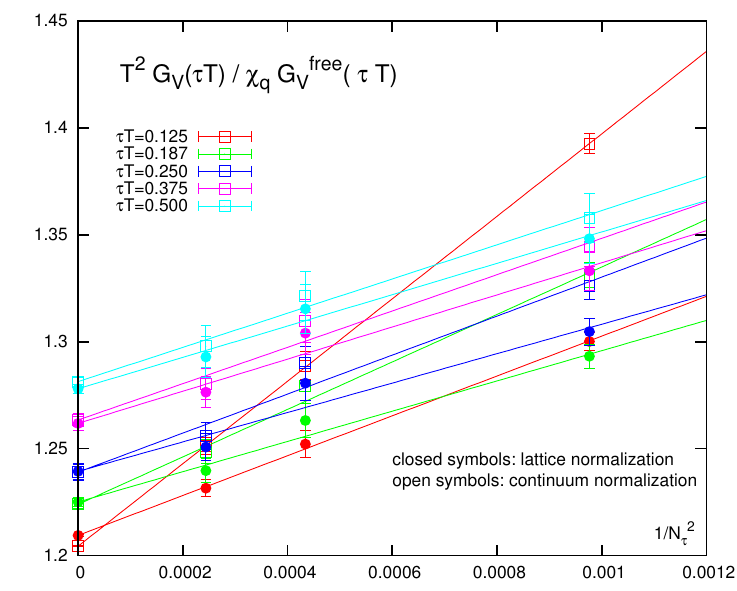}
\includegraphics{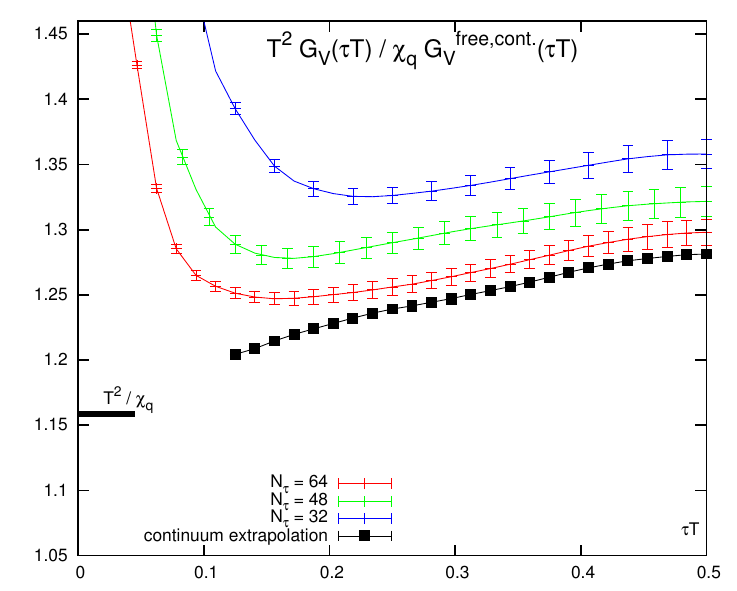}
\caption{\textit{Left:} Continuum 
extrapolation of the correlator ratio 
$G_V(\tau T)/G_V^\text{free,lat.}(\tau T)$ and 
$G_V(\tau T)/G_V^\text{free,cont.}(\tau T)$
at different time-like separations $\tau T$. Cutoff
effects are under control for $0.125 < \tau T < 0.5$.
\textit{Right:} Correlator ratio 
$G_V(\tau T)/G_V^\text{free,cont.}(\tau T)$ for the
three lattice spacings and continuum extrapolation.}
\label{corr-extr}
\end{figure}

Cut-off effects are removed by a 
continuum extrapolation of the correlators: 
Discretization errors of non-perturbatively 
clover-improved Wilson fermions 
have a quadratic error in the lattice spacing
allowing to extrapolate the correlators in $a^2$, 
corresponding to $1/N^2_\tau$ at fixed temperature $T$.
For these extrapolations, the ratios \eqref{fitRatios} 
of the interacting correlators to free theory are used. 
On the coarser lattices these ratios are spline-interpolated to
provide every spacing $\tau T$ present in the finest correlator.
As can be seen in figure \ref{corr-extr}, calculating ratios to
the free continuum as well as to the free lattice correlators
give the same continuum extrapolation for Euclidean time
separations $\tau T > 0.125 $ and the results show that
lattice cutoff effects are under control 
in the region $0.15 < \tau T < 0.5$.

\subsection{Fitting the spectral function}

\begin{figure}
\includegraphics{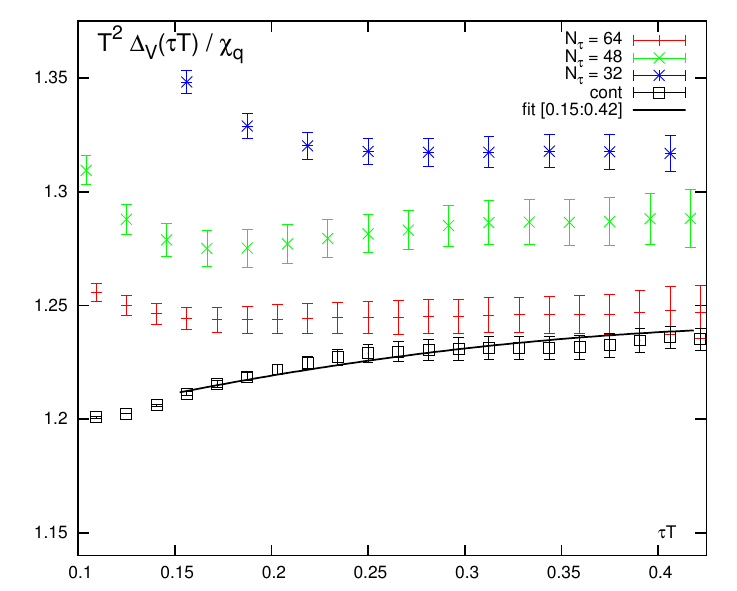}
\includegraphics{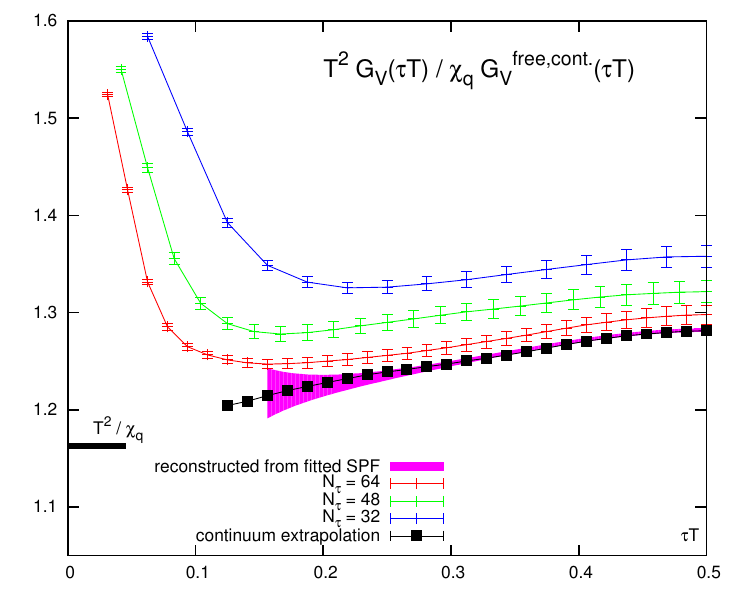}
\caption{\textit{Left:} 
Ratio of midpoint substracted correlators 
$  ( G_V(\tau T) - G^{(0)}_V ) / ( G_V^\text{free}(\tau T) - G^{(0),\text{free}}_V ) $, including a fit of 
the curvature \protect\eqref{midpoint-sub-corr} and thereby the thermal moments.
\textit{Right:} Fit of the continuum extrapolated correlator to \protect\eqref{spfToCorr} 
with the spectral function ansatz \protect\eqref{spfAnsatz}.
}
\label{midpoint}
\end{figure}

After the continuum extrapolation of the vector correlation
function has been obtained, the first and second 
thermal moments are computed
by fitting the curvature of the midpoint subtracted correlator
\eqref{midpoint-sub-corr}.
Both the correlator ratio 
$G_V(\tau T)/\chi_q G_V^\text{free}(\tau T)$ 
and the ratio of the first thermal moments serve as input and 
constraints for a fit of the parameters $c_{BW}, \Gamma, \kappa$ 
in the spectral 
function ansatz \eqref{spfAnsatz} to reproduce the correlator 
\eqref{spfToCorr}. As can be seen in figure \ref{midpoint} (right), the
fitted spectral function reproduces the correlation functions. 
The thermal moments constrain the small $\omega$ region, so
errors of the spectral function fit descrease with increasing $\tau T$.

\subsection{Breit-Wigner with truncated continuum}

As in the analysis of the $1.45 T_c$ dataset \cite{Ding:2010ga}, 
to study the systematic uncertainties of our Ansatz for the spectral function
the low energy structure of the spectral function
was studied by smoothly cutting the continuum contribution
at a frequency $\omega_0$ with a width of $\Delta_\omega$

\begin{align}
&\rho_{ii}^\text{trunc.}(\omega) = \chi_q c_{\text{BW}} \frac{\omega \Gamma}{\omega^2 + (\Gamma/2)^2}
 + (1+\kappa) \frac{3}{2\pi} \omega^2 \tanh(\omega/4T) 
 \Theta(\omega_0, \Delta_\omega)  
\label{spfTrunc} \\
&\text{with} \quad
\Theta(\omega_0, \Delta_\omega) = \left ( 1 + e^{(\omega_0^2 - \omega^2)/\omega \Delta_\omega} \right )^{-1} \notag
\end{align}

\begin{figure}
\includegraphics{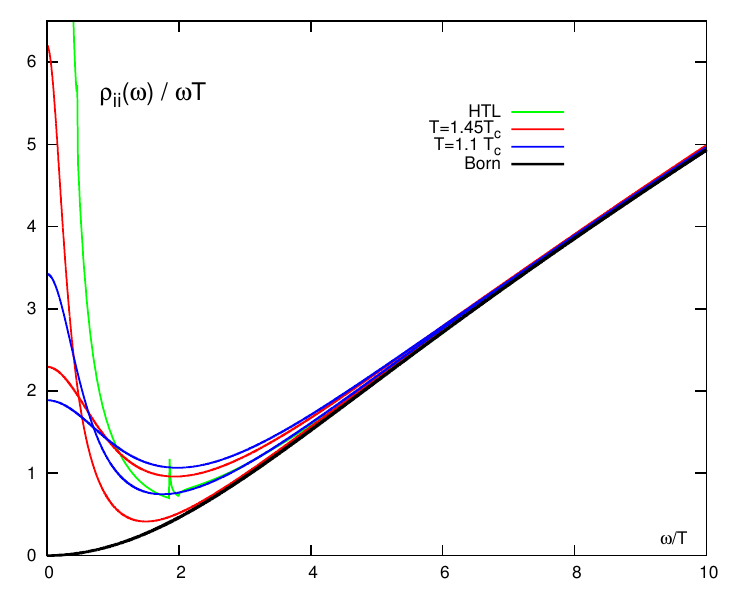}
\includegraphics{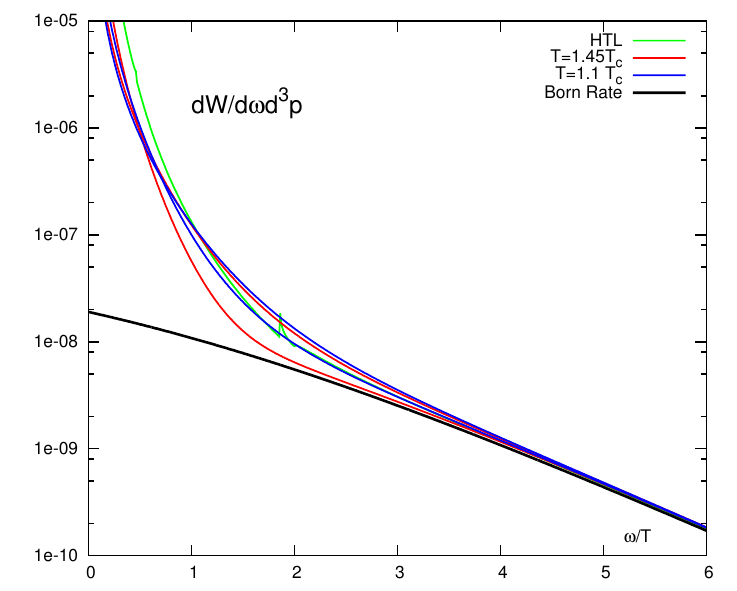}
\caption{\textit{Left:} Spectral function as fitted to the 
vector correlation function. 
Two lines are shown for each temperature as a result
of the systematic error estimation,
given by the two $\rho_{ii}$
with the minimal and maximal
$c_\text{BW}$ that can be obtaind 
via the Breit-Wigner + truncated continuum ansatz \protect\eqref{spfTrunc},
while maintaining a $\chi/\text{d.o.f} < 1.1$ in the fit.
Results from
the hard thermal loop resummation scheme (HTL) are also
included in the plot \protect\cite{Braaten:1989mz}.
\textit{Right:} Thermal dilepton
rate calculated from the spectral function \protect\eqref{dileptonrate}.
}
\label{spfAndDil}
\end{figure}

The fit with \eqref{spfTrunc} is performed for
a range of values $\omega_0$ and $\Delta_\omega$. 
The systematic errors given in figure \ref{spfAndDil}
correspond to the minimal and maximal 
values $c_{\text{BW}}$ found in these fits with
$\chi^2/\text{d.o.f} < 1.1 $, thereby giving
a minimal and maximal electrical conductivity \eqref{elcon}
within this framework. 

From the corresponding spectral functions 
the dilepton rates shown in
figure \ref{spfAndDil} (right) are calculated \eqref{dileptonrate}.
Within the current systematic uncertainties, 
the dilepton rates 
and electrical conductivities in units of $T$
agree between 
$1.1 T_c$ and $1.45 T_c$.
This linear temperature dependence of the electrical conductivity 
is in agreement with the results of
\cite{Aarts:2007wj}.

In order to study systematic errors,
also a variation of the ansatz \eqref{spfTrunc}
motivated by Operator Product Expansion
\cite{Burnier:2012ts} can be used. First tests
show only small deviations that are well 
within the systematic errors between
both version of the ansatz.

\section{Outlook: Correlation functions at finite momenta}

\begin{figure}
\includegraphics{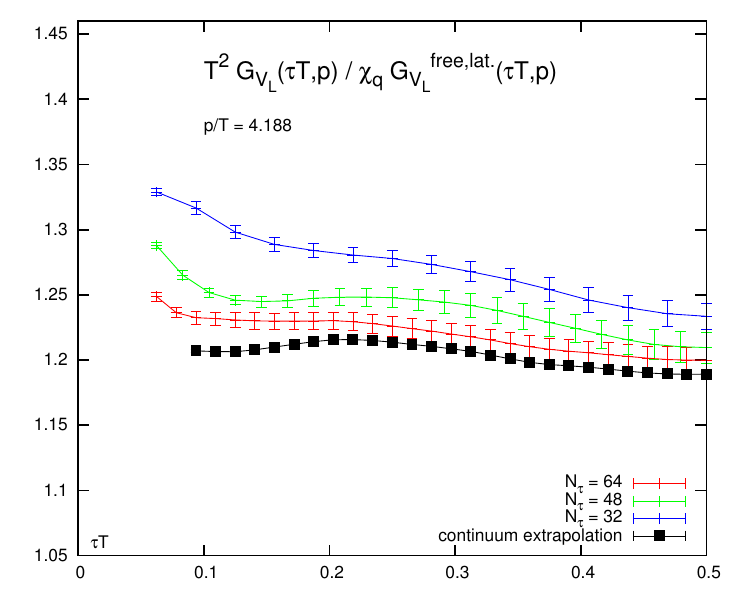}
\includegraphics{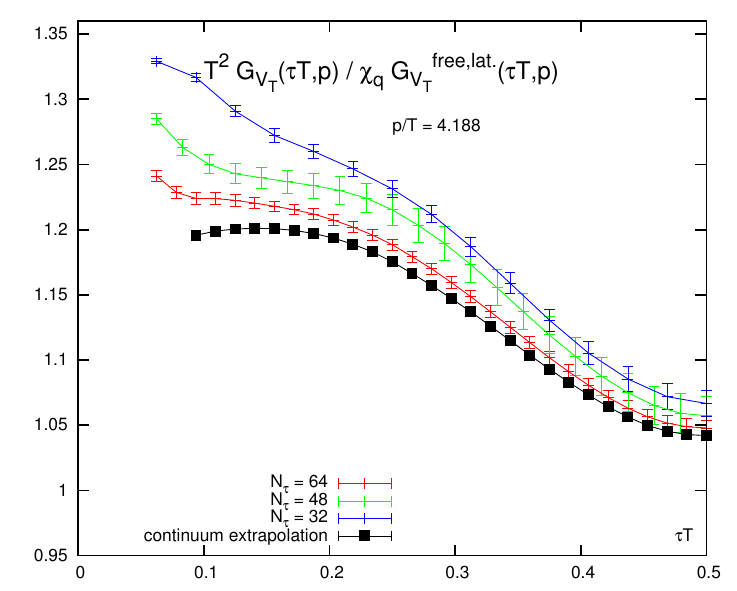}
\caption{Continuum extrapolation of the 
logitudinal (left) and transversal (right) 
vector correlation function
$G_V/G_V^\text{free}$ at non-zero momentum.}
\label{finiteMomentum}
\end{figure}

Extracting the spectral function at vanishing momentum $\vec p=0$
gives access to the dilepton rates and the electric conductivity.
For finite momenta $\vec p \neq 0$ the
correlation and spectral functions split into a longitudinal
and transversal part, where
the transversal 
$\rho_T$ relates
to the photon rate:

\begin{equation}
\omega \frac{d R_\gamma}{d^3p} \sim  
\frac{\rho_T(\omega=|\vec{p}|,T)}{\exp(\omega/T)-1}.
\end{equation}

On the lattice, only a finte number of discrete momenta are accessible,
the momenta are given by the aspect ratio $N_\sigma / N_\tau$ and
a vector of integer numbers $\vec k$ as
$\vec{p}/T = 2 \pi \cdot \vec{k} \cdot N_\tau / N_\sigma$. All
$1.1 T_c$ correlator calculations have been performed on the
same aspect ratio of $N_\sigma / N_\tau = 3$, so a continuum
extrapolation of the correlator at fixed momenta can be performed.
As can be seen in figure \ref{finiteMomentum}, this extrapolation
is well behaved as in the zero momentum case and gives precise
results for the vector correlation function at finite momentum 
$\vec p$ in the continuum limit. Work on the extraction of the
corresponding spectral functions and the determination 
of the photon rate is in progress.

\section{Conclusion}

Lattice calculations of the vector correlation function
have been performed for two temperatures in the
deconfined phase of QCD. Calculations at different
lattice spacings were used for a successfull continuum
extrapolation to remove cut-off effects. Using a
phenomenologically motivated ansatz, the vector
spectral function to the continuum-extrapolated correlation
function was obtained, giving access to the dilepton
rate and the electric conductivity of the medium at
the given temperatures. 
Within current systematic uncertainties, 
the electric conductivity divided by temperature 
and the thermal dilepton rates are
are compatible at $1.1 T_c$ and $1.45 T_c$.
A continuum extrapolation was
also performed for the correlation functions at
a range of finite momenta, opening the
possibility to study the photon rate in the future.

\section{Acknowlegdements}

The results for the vector correlation functions have been 
achieved by using the PRACE Research 
Infrastructure resource JUGENE based at the 
J\"ulich Supercomputing Centre in Germany
and GPU-cluster resources of the 
lattice gauge theory group at Bielefeld University.
This work is supported by the 
IRTG/GRK 881 "Quantum Fields and Strongly Interacting Matter".

\bibliographystyle{JHEP}
\bibliography{literature}

\providecommand{\href}[2]{#2}\begingroup\raggedright\begin{thebibliography}{10}

\bibitem{Rapp:2009yu}
R.~Rapp, J.~Wambach, and H.~van Hees, {\it {The Chiral Restoration Transition
  of QCD and Low Mass Dileptons}},  in {\em Landolt-B{\"o}rnstein}, vol.~I-23,
  4-1.
\newblock Springer-Verlag, 2010.
\newblock \href{http://arxiv.org/abs/0901.3289}{{\tt arXiv:0901.3289}}.

\bibitem{pos-olaf}
O.~Kaczmarek, {\it {Recent Developments in Lattice Studies for Quarkonia}},
  {\em to appear in Nucl. Phys. A} [\href{http://arxiv.org/abs/1208.4075}{{\tt
  arXiv:1208.4075}}].

\bibitem{pos-hengtong}
H.-T. Ding, {\it In-medium hadron properties from lattice qcd},  {\em EPJ Web
  of Conferences} {\bf 36} (2012) 00008,
  [\href{http://arxiv.org/abs/1207.5187}{{\tt arXiv:1207.5187}}].

\bibitem{Ding:2010ga}
H.-T. Ding, A.~Francis, O.~Kaczmarek, F.~Karsch, E.~Laermann, et~al., {\it
  {Thermal dilepton rate and electrical conductivity: An analysis of vector
  current correlation functions in quenched lattice QCD}},  {\em Phys.Rev.}
  {\bf D83} (2011) 034504, [\href{http://arxiv.org/abs/1012.4963}{{\tt
  arXiv:1012.4963}}].

\bibitem{Francis:2011bt}
A.~Francis and O.~Kaczmarek, {\it {On the temperature dependence of the
  electrical conductivity in hot quenched lattice QCD}},  {\em
  Prog.Part.Nucl.Phys.} {\bf 67} (2012) 212--217,
  [\href{http://arxiv.org/abs/1112.4802}{{\tt arXiv:1112.4802}}].

\bibitem{Brandt:2012jc}
B.~B. Brandt, A.~Francis, H.~B. Meyer, and H.~Wittig, {\it {Thermal Correlators
  in the $\rho$ channel of two-flavor QCD}},
  \href{http://arxiv.org/abs/1212.4200}{{\tt arXiv:1212.4200}}.

\bibitem{Allton:2005gk}
C.~Allton, M.~Doring, S.~Ejiri, S.~Hands, O.~Kaczmarek, et~al., {\it
  {Thermodynamics of two flavor QCD to sixth order in quark chemical
  potential}},  {\em Phys.Rev.} {\bf D71} (2005) 054508,
  [\href{http://arxiv.org/abs/hep-lat/0501030}{{\tt hep-lat/0501030}}].

\bibitem{Gavai:2001ie}
R.~V. Gavai, S.~Gupta, and P.~Majumdar, {\it {Susceptibilities and screening
  masses in two flavor QCD}},  {\em Phys.Rev.} {\bf D65} (2002) 054506,
  [\href{http://arxiv.org/abs/hep-lat/0110032}{{\tt hep-lat/0110032}}].

\bibitem{Aarts:2002cc}
G.~Aarts and J.~M. Martinez~Resco, {\it {Transport coefficients, spectral
  functions and the lattice}},  {\em JHEP} {\bf 0204} (2002) 053,
  [\href{http://arxiv.org/abs/hep-ph/0203177}{{\tt hep-ph/0203177}}].

\bibitem{Moore:2006qn}
G.~D. Moore and J.-M. Robert, {\it {Dileptons, spectral weights, and
  conductivity in the quark-gluon plasma}},
  \href{http://arxiv.org/abs/hep-ph/0607172}{{\tt hep-ph/0607172}}.

\bibitem{Petreczky:2005nh}
P.~Petreczky and D.~Teaney, {\it {Heavy quark diffusion from the lattice}},
  {\em Phys.Rev.} {\bf D73} (2006) 014508,
  [\href{http://arxiv.org/abs/hep-ph/0507318}{{\tt hep-ph/0507318}}].

\bibitem{Hong:2010at}
J.~Hong and D.~Teaney, {\it {Spectral densities for hot QCD plasmas in a
  leading log approximation}},  {\em Phys.Rev.} {\bf C82} (2010) 044908,
  [\href{http://arxiv.org/abs/1003.0699}{{\tt arXiv:1003.0699}}].

\bibitem{Aarts:2007wj}
G.~Aarts, C.~Allton, J.~Foley, S.~Hands, and S.~Kim, {\it {Spectral functions
  at small energies and the electrical conductivity in hot, quenched lattice
  QCD}},  {\em Phys.Rev.Lett.} {\bf 99} (2007) 022002,
  [\href{http://arxiv.org/abs/hep-lat/0703008}{{\tt hep-lat/0703008}}].

\bibitem{Burnier:2012ts}
Y.~Burnier and M.~Laine, {\it {Towards flavour diffusion coefficient and
  electrical conductivity without ultraviolet contamination}},  {\em
  Eur.Phys.J.} {\bf C72} (2012) 1902,
  [\href{http://arxiv.org/abs/1201.1994}{{\tt arXiv:1201.1994}}].

\bibitem{Braaten:1989mz}
E.~Braaten and R.~D. Pisarski, {\it {Soft Amplitudes in Hot Gauge Theories: A
  General Analysis}},  {\em Nucl.Phys.} {\bf B337} (1990) 569.

\end{thebibliography}\endgroup

\end{document}